\documentclass[aps,prl,twocolumn,groupedaddress]{revtex4-1}
\usepackage{amsmath}
\usepackage{graphicx}

\usepackage{xparse}
\NewDocumentCommand{\rot}{O{45} O{1em} m}{\makebox[#2][l]{\rotatebox{#1}{#3}}}%

\usepackage{hyperref}

\begin{document}

% Use the \preprint command to place your local institutional report
% number in the upper righthand corner of the title page in preprint mode.
% Multiple \preprint commands are allowed.
% Use the 'preprintnumbers' class option to override journal defaults
% to display numbers if necessary
\preprint{MITP/19-033}

%Title of paper
\title{Two- and three-pion finite-volume spectra at maximal isospin from lattice QCD}

% repeat the \author .. \affiliation  etc. as needed
% \email, \thanks, \homepage, \altaffiliation all apply to the current
% author. Explanatory text should go in the []'s, actual e-mail
% address or url should go in the {}'s for \email and \homepage.
% Please use the appropriate macro foreach each type of information

% \affiliation command applies to all authors since the last
% \affiliation command. The \affiliation command should follow the
% other information
% \affiliation can be followed by \email, \homepage, \thanks as well.
\author{Ben~H\"{o}rz}
\email{hoerz@lbl.gov}
\affiliation{Nuclear Science Division, Lawrence Berkeley National Laboratory,
Berkeley, CA 94720, USA}
\author{Andrew~Hanlon}
\email{ahanlon@uni-mainz.de}
\affiliation{Helmholtz-Institut Mainz, Johannes Gutenberg-Universit\"{a}t,
55099 Mainz, Germany}

\date{\today}

\begin{abstract}
We present the three-pion spectrum with maximal isospin in a finite volume
determined from lattice QCD,
including excited states in addition to the ground states
across various irreducible representations at zero and nonzero total momentum.
The required correlation functions, from which the spectrum is extracted, are
computed using a newly implemented algorithm which speeds up the computation
by more than an order of magnitude.
On a subset of the data we extract a nonzero value of the three-pion
threshold scattering amplitude using the $1/L$ expansion of the three-particle
quantization condition, which consistently describes all states
at zero total momentum.
The finite-volume spectrum is
publicly available to facilitate further explorations within
the available three-particle finite-volume approaches.
\end{abstract}

% insert suggested PACS numbers in braces on next line
% insert suggested keywords - APS authors don't need to do this
%\keywords{}

%\maketitle must follow title, authors, abstract, \pacs, and \keywords
\maketitle

% body of paper here - Use proper section commands
% References should be done using the \cite, \ref, and \label commands
\section{Introduction\label{s:intro}}
% Put \label in argument of \section for cross-referencing
%\section{\label{}}
Lattice QCD calculations of scattering amplitudes have matured significantly 
over the last decade owing to marked increases in available computational
capacity and improved algorithms.
A widely used approach for constraining scattering observables
from simulations relies on precise measurements of the
interacting energy levels of QCD in a finite volume, which encode
hadron interactions via the shifts from their noninteracting values~\cite{
Luscher:1985dn,Luscher:1986pf,Luscher:1990ux,Luscher:1990ck,Rummukainen:1995vs} 
(see \cite{Briceno:2017max} for a survey of extensions of the
formalism and numerical results).

So far, practical calculations in lattice QCD have been mostly confined to the
two-hadron sector. 
Though a large abundance of lattice data is currently
available for scattering of two hadrons
(e.g. $\pi \pi$ scattering in all three isospin channels
\cite{Beane:2005rj,Beane:2007xs,Feng:2009ij,Feng:2010es,Beane:2011sc,Lang:2011mn,Aoki:2011yj,Pelissier:2012pi,Dudek:2012gj,Dudek:2012xn,Sasaki:2013vxa,Fu:2013ffa,Wilson:2015dqa,Helmes:2015gla,Bali:2015gji,Bulava:2016mks,Guo:2016zos,Fu:2016itp,Briceno:2016mjc,Liu:2016cba,Alexandrou:2017mpi,Briceno:2017qmb,Andersen:2018mau,Guo:2018zss,Culver:2019qtx},
see also \cite{Kurth:2013tua,Akahoshi:2019klc} for results
using a potential-based approach),
these calculations are formally restricted to energies below
thresholds involving three or more hadrons
due to the use of a formalism for relating finite-volume spectra to
scattering amplitudes that is limited to two-hadron scattering.
This limitation has precluded a proper lattice
QCD study of systems involving three or more stable
hadrons at light pion masses,~e.g. the Roper resonance
which decays to both two- and three-particle channels,
the $\omega(782)$ decaying to three pions, many of the
$X$, $Y$ and $Z$ resonances, and three-nucleon interactions relevant
for nuclear physics.

Following the demonstration that the finite-volume spectrum is determined by the
infinite-volume $S$ matrix, even in the presence of three-particle
intermediate states \cite{Polejaeva:2012ut},
significant progress has been made in developing the necessary
formalism to interpret the three-particle finite-volume spectrum, both by
extending the two-particle derivation to include three-hadron states
\cite{Hansen:2014eka,Hansen:2015zga,Briceno:2017tce,Briceno:2018aml},
as well as through alternative approaches \cite{Agadjanov:2016mao,Hammer:2017uqm,Hammer:2017kms,Mai:2017bge,Doring:2018xxx}
(for a review see \cite{Hansen:2019nir}).\footnote{See also
\cite{Hashimoto:2017wqo,Hansen:2017mnd,Bulava:2019kbi}
for a complementary approach for inclusive observables.}
Thus, although the three-particle formalism is quite mature\textemdash
including numerical explorations of the corresponding quantization conditions
\cite{Briceno:2018mlh,Romero-Lopez:2018rcb,Mai:2018djl}%
\textemdash data for three-particle
finite-volume QCD spectra
is lacking, since previous lattice QCD calculations have
been restricted to the extraction of multi-meson ground states at rest \cite{Beane:2007es,Detmold:2008fn,Detmold:2008yn}.
Recently a first calculation of finite-volume spectra including
three-particle energies was carried out in the $b_1$
system, whose results are however not yet
amenable to an interpretation in the
present three-particle formalisms \cite{Woss:2019hse}.
Hence no comprehensive data exists to apply the available finite-volume
formalisms.

We fill this gap by providing the two-pion and three-pion spectra with
maximum isospin in various irreducible representations (irreps) at zero and
nonzero total momentum in the elastic region, i.e. for center-of-mass energies
$E_\mathrm{cm}/m_\pi$ below 4 and 5 for isospin $I=2$ and $I=3$ respectively.
Our analysis of a subset of the data indicates sensitivity to the three-pion
interaction at the current level of precision.
In order to facilitate a more detailed exploration, possibly
including the effect of higher partial
waves \cite{Blanton:2019igq,Hansen:2014eka,Hansen:2015zga,Doring:2018xxx},
the spectrum data is made public including all correlations.

A technical challenge concerns the growing number of Wick contractions
required to compute correlation functions of suitable interpolating operators%
\textemdash from which the spectrum is extracted\textemdash%
as the number of valence quark fields increases.
The continued need for improved algorithms to perform these contractions
was pointed
out recently \cite{Detmold:2019ghl} and indeed was a limiting factor in
a recent study of meson-baryon scattering in the $\Delta$
channel \cite{Andersen:2017una}.
While Refs. \cite{Yamazaki:2009ua,Detmold:2010au,Doi:2012xd,Detmold:2012eu,Gunther:2013xj,Nemura:2015yha} investigated
efficient contraction algorithms at the quark level,
we employ the stochastic
variant \cite{Morningstar:2011ka} of distillation \cite{Peardon:2009gh} to
treat quark propagation.
In this framework, it is useful to view the correlation function
construction in terms of contractions of tensors associated
with the involved hadrons.
Then, to reduce the operation count required to evaluate all contractions, we use a method
which is well-known in quantum
chemistry \cite{05-TR10,10.1007/11758501_39,doi:10.1021/jp9051215} and
has attracted renewed interest in the context of
tensor networks \cite{PhysRevE.90.033315}.
The proposed optimization achieves a speedup by
more than an order of magnitude,
can be readily used for general physical
systems (e.g. three-meson systems at nonmaximal isospin and two-baryon systems),
and its implementation is publicly available
\footnote{\url{https://github.com/laphnn/contraction_optimizer}}.

This letter is organized as follows: We first discuss the interpolating
operators employed and construction of
their correlation functions,
followed by a description of the ensemble used in this work.
Subsequently the finite-volume spectra and extraction of two- and three-pion
scattering parameters from a subset of the data are presented.

\section{\label{s:meth}Lattice QCD methods}
\emph{Interpolating operators}: Lattice simulations of QCD in a cubic box
break the infinite-volume $SO(3)$ rotational symmetry.
The spectrum in finite
volume is customarily extracted employing interpolating operators which
transform irreducibly under the symmetries of a cubic spatial lattice,
%(see e.g. \cite{Basak:2005aq,Basak:2005ir} for baryon operators),
i.e. the octahedral group $O_h$ for zero total momentum
$\textbf{P} = \textbf{0}$ and the corresponding little groups
for $\textbf{P} \neq \textbf{0}$ \cite{Dudek:2012gj,Morningstar:2013bda}.
Correlation functions of such interpolators access only the
sub-block of the finite-volume Hamiltonian corresponding to the same irrep,
thus greatly simplifying the determination of
the spectrum and the subsequent scattering-amplitude analysis.

We employ the simplest single-pion operator destroying a
three-momentum $\textbf{p}$ given by
\begin{align}
	\pi_{\textbf{p}}(t) = \sum_{\textbf{x}} \mathrm{e}^{-i \textbf{p}\cdot\textbf{x}} \bar d(\textbf{x},t) \gamma_5 u(\textbf{x},t).
	\label{eqn:singlepi}
\end{align}
This operator transforms in the $A_{1u}^-$ and $A_{2}^-$ irrep
for zero and nonzero momentum respectively, where the superscript
specifies the $G$-parity.

Two-pion interpolators which transform according to the irrep $\Lambda$ of
the little group of total momentum $\textbf{P}$ are obtained by
forming appropriate linear combinations of two single-pion interpolators
with momenta $\textbf{p}_1 + \textbf{p}_2 = \textbf{P}$,
\begin{align}
	\pi \pi^{(\textbf{P}, \Lambda)}(t) = c^{(\textbf{P},\Lambda)}_{\textbf{p}_1,\textbf{p}_2} \pi_{\textbf{p}_1}(t) \pi_{\textbf{p}_2}(t).
	\label{eqn:twopi}
\end{align}
The relevant Clebsch-Gordan coefficients
$c^{(\textbf{P},\Lambda)}_{\textbf{p}_1,\textbf{p}_2}$
were worked out in \cite{Morningstar:2013bda} (see also \cite{Dudek:2012gj})
and used previously to study $\pi\pi$
scattering \cite{Bulava:2016mks,Andersen:2018mau}.

Three-pion interpolators are obtained by iterating this process, i.e. by
first coupling two of the pions into an intermediate irrep,
then using the Clebsch-Gordan coefficients again to obtain operators
transforming according to one of the total irreps of interest,
\begin{align}
	\pi \pi \pi^{(\textbf{P}, \Lambda)}(t) = c^{(\textbf{P},\Lambda)}_{\textbf{p}_1,\textbf{p}_2,\textbf{p}_3} \pi_{\textbf{p}_1}(t) \pi_{\textbf{p}_2}(t) \pi_{\textbf{p}_3}(t).
	\label{eqn:threepi}
\end{align}
Due to the weak interaction in $I=2$ $\pi\pi$ scattering, which is
the only relevant subprocess for this work, the more elaborate operator
construction discussed in \cite{Woss:2019hse} is not required.
The interpolators we use in this work are listed in the supplemental material.

\begin{figure}
\includegraphics{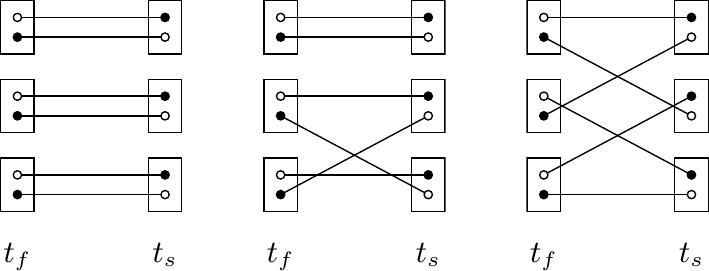}%
\caption{\label{fig:3picontracts}Different topologies of Wick contractions
required to evaluate $I=3$ three-pion correlation functions.
Circles indicate quark and antiquark fields tied into meson functions shown
as boxes, which are subsequently contracted.
Two-pion $I=2$ Wick contractions appear as subexpressions.}
\end{figure}
\emph{Correlation function construction}: Quark propagation is
treated using the stochastic LapH method
\cite{Morningstar:2011ka}
by first obtaining smeared solutions of the Dirac equation,
\begin{align}
  \varphi^{[r,d]} = \mathcal{S} D^{-1} \varrho^{[r,d]} ,
\end{align}
for stochastic quark-field sources $\varrho^{[r,d]}$ with
noise index $r=1,\dots,N_\eta$, dilution \cite{Wilcox:1999ab,Foley:2005ac}
index $d=1,\dots,N_\mathrm{dil}$,
and where $\mathcal{S}$ is the LapH smearing kernel, formed
from the $N_\mathrm{ev}$ lowest eigenvectors of the three-dimensional covariant Laplacian.
Next, useful intermediate quantities
are the pion source and sink functions
%\footnote{We have used $\gamma_5$-hermiticity
%of the quark propagator in the definition of the pion function
%which cancels the $\gamma_5$ of the single-pion interpolator
%in Eq. \eqref{eqn:singlepi}.}
\cite{Morningstar:2011ka}
\begin{equation}
\begin{aligned}
	\mathcal{M}_{\textbf{p}}^{[r_1,r_2,d_1,d_2]}(t) &= -
		\sum_{\textbf{x}} \mathrm{e}^{-i \textbf{p}\cdot\textbf{x}}
		\varrho^{[r_1,d_1]*}_{a\alpha \textbf{x} t}
		\varrho^{[r_2,d_2]}_{a\alpha \textbf{x} t}, \\
		\bar{\mathcal{M}}_{\textbf{p}}^{[r_1,r_2,d_1,d_2]}(t) &= \phantom{-}
		\sum_{\textbf{x}} \mathrm{e}^{-i \textbf{p}\cdot\textbf{x}}
		\varphi^{[r_1,d_1]*}_{a\alpha \textbf{x} t}
		\varphi^{[r_2,d_2]}_{a\alpha \textbf{x} t},
\end{aligned}
\label{eqn:mesfunc}
\end{equation}
with summed color index $a$ and spin index $\alpha$, and two open
noise and dilution indices.
In terms of these meson functions, the single-pion correlation function
on a single gauge configuration is obtained by the average
over noise combinations $\{r_1,r_2\}$ \cite{Morningstar:2011ka}
\begin{align}
	C_{\pi_\textbf{p}}(t_f&-t_s) \propto \nonumber \\& -\sum_{\{r_1,r_2\}}
		\bar{\mathcal{M}}_{\textbf{p}}^{[r_1,r_2,d_1,d_2]}(t_f)
		\mathcal{M}_{\textbf{p}}^{[r_1,r_2,d_1,d_2]*}(t_s),
	\label{eqn:picorr}
\end{align}
with proper normalization given by the number of noise
combinations used to perform the average.

For a given momentum, pair of source and sink time $t_s$ and $t_f$, and noise combination,
Eq. \eqref{eqn:picorr} is a tensor contraction over dilution
indices of two rank-2 tensors with index range $N_\mathrm{dil}$.
Two- and three-pion correlation functions with maximal isospin can
be computed using the same building blocks \cite{Morningstar:2011ka}
and involve tensor contractions governed by all possible Wick contractions
of four and six rank-2 tensors respectively.

The number of Wick contractions grows factorially as more pions are included
\cite{Beane:2007es}, and the different topologies of diagrams for three-pion
correlation functions in the sector of maximal isospin are shown
in Fig. \ref{fig:3picontracts}.
However, across the relevant diagrams there is a lot of redundancy which
can be exploited systematically to reduce the number of arithmetic operations
required for their evaluation.
In particular, all diagrams required for the computation of
$I=2$ two-pion correlation functions appear as subdiagrams of
$I=3$ three-pion correlation functions.

The algorithm to automatically perform the operation-count minimization is
described in the supplemental material together with a detailed example.
For the evaluation of the correlation functions required in this work we
achieve a speed-up by roughly a factor of 15.

\begin{table}
\caption{\label{tab:ens}Ensemble and measurement setup used in this work.
Measurements are performed on $N_\mathrm{cfg}$ configurations
with spatial volume $L^3$
separated by $4 \, \mathrm{MDU}$.
Correlation functions are estimated starting from a single source
time $t_s/a = 35$ using $N_\eta$ diluted noise sources to estimate quark
propagators (see \cite{Morningstar:2011ka} for unexplained notation).
Other parameters, e.g. for stout smearing \cite{Morningstar:2003gk}, are the same as in \cite{Andersen:2018mau}.}
\begin{ruledtabular}
\begin{tabular}{c c c c c c}
	$a m_\pi$ & $L/a$ &  $N_\mathrm{ev}$ & dilution & $N_\eta$ & $N_\mathrm{cfg}$ \\
	\hline
	0.06504(33) & 64 & 448 & (TF,SF,LI16) & 6 & 1100 \\
\end{tabular}
\end{ruledtabular}
\end{table}
\emph{Ensemble details}: The results in this work are based on the
D200 ensemble generated through the CLS effort \cite{Bruno:2014jqa}
with $N_\mathrm{f}=2+1$ quark flavors at a pion mass
$m_\pi \approx 200 \, \mathrm{MeV}$ and lattice spacing $a \approx 0.064 \, \mathrm{fm}$ \cite{Bruno:2016plf}.
%The simulation belongs to a chiral trajectory where $2 m_l + m_s$ is
%kept fixed as the light quark mass $m_l = m_u = m_d$ is lowered
%towards its physical value, implying $m_\pi > m_\pi^\mathrm{phys}$
%and $m_K < m_K^\mathrm{phys}$.
The ensemble and measurement setup is detailed in Table \ref{tab:ens}.
In order to ensure a Hermitian matrix of correlation functions
despite the use of open boundary conditions in the temporal direction \cite{Luscher:2011kk},
our interpolating operators are always separated from the temporal boundaries
by at least $t_s$, where $m_\pi t_s = 2.2$.

Pion-pion scattering in the isovector channel has been investigated
on this ensemble previously \cite{Andersen:2018mau}, and statistics subsequently
improved considerably to provide spectroscopic information for the
determination of the hadronic vacuum polarization on the same
ensemble \cite{Gerardin:2019rua}.
The pion functions are re-used from that work, and hence no
additional meson functions or solutions of the Dirac equation have to be
computed.

\section{Results\label{s:res}}
\emph{Analysis strategy}: The procedure to extract the finite-volume spectrum
from a matrix of correlation functions $C_{ij}(t)$ in a given irrep
is discussed in detail in \cite{Bulava:2016mks} and we use the
analysis suite \footnote{\url{https://github.com/ebatz/jupan}} developed in \cite{Andersen:2018mau}.

We solve a generalized eigenvalue problem
\cite{Michael:1982gb,Luscher:1990ck,Blossier:2009kd}
for a fixed reference time and diagonalization time $(t_0=5a, t_*=10a)$,
corresponding to roughly $0.32 \, \mathrm{fm}$ and
$0.64 \, \mathrm{fm}$ in physical units \cite{Bruno:2016plf},
in order to extract not only the ground state but also excited states in most irreps.
Results from different $(t_0, t_*)$ are indistinguishable,
presumably
due to the weak interaction in $I=2$ and $I=3$ pion scattering which results
in little mixing of our interpolating operators,
in which each hadron has been projected to definite momentum and
is hence expected to overlap predominantly with a single state.

For two-pion states the difference $\Delta \! E$ between interacting
and noninteracting energies is determined from single-exponential
fits at sufficiently large time separations to the ratios
\begin{align}
	R_i(t) = \frac{\hat{C}_{ii}(t)}{C_{\pi_{\textbf{p}_1}}\!(t)
	\, C_{\pi_{\textbf{p}_2}}\!(t)} \xrightarrow{\mathrm{large}\; t}
	A \mathrm{e}^{-\Delta \! E_i t}
\end{align}
of diagonal elements of the `optimized' correlation
matrix $\hat C$ \cite{Bulava:2016mks}
%(i.e. the matrix formed from rotations by the eigenvectors of the generalized eigenvalue problem)
and two single-pion correlation functions, and
similarly for the three-pion states \cite{Beane:2007es}.
Absolute energies are reconstructed from those energy differences using
the single-pion dispersion relation.
%The two- and three-pion spectra with maximum isospin are extracted across
%a number of irreps with zero and nonzero total momentum.
The attainable precision is generally at
the few-permille level for the energies measured in units of the single-pion
mass.

\begin{figure}
\includegraphics{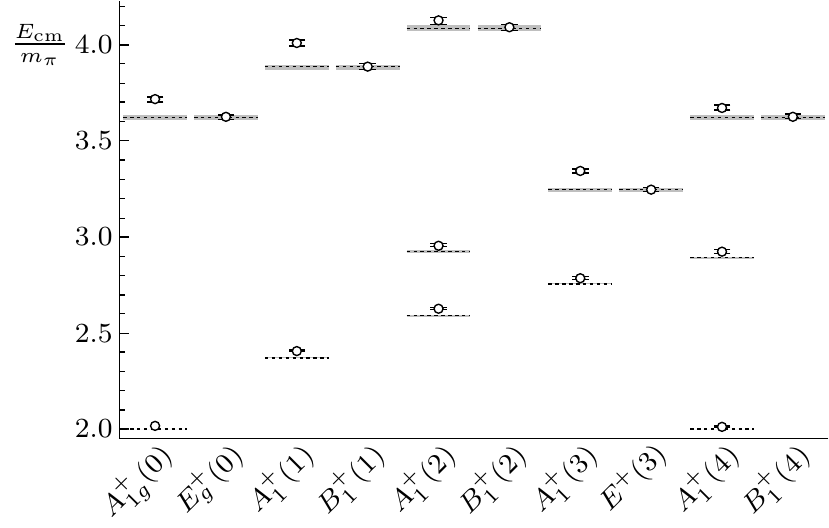}%
\caption{\label{fig:res2pi}$I=2$ two-pion spectrum in various irreps
$\Lambda(\textbf{d}^2)$ with total momentum
$\textbf{P} = \frac{2\pi}{L} \textbf{d}$.
Open symbols denote the measured interacting energies which are shifted
from their noninteracting values shown as dashed lines.}
\end{figure}
\begin{figure}
\includegraphics{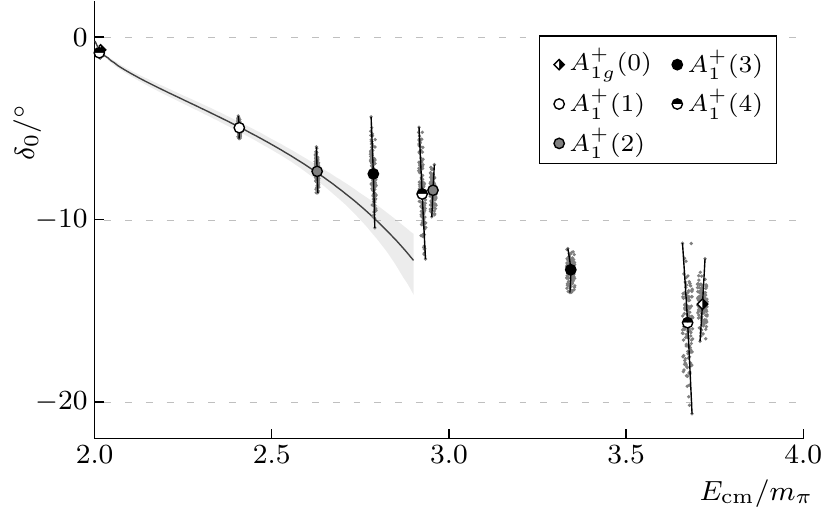}%
\caption{\label{fig:phase_shift} Energy dependence of the $I=2$ $\pi \pi$
s-wave scattering amplitude extracted from the two-pion spectrum.
The gray band shows the result of the fit to the five leftmost points using
the effective range expansion given in Eq. \eqref{eqn:ere}.}
\end{figure}
\emph{Two-pion spectrum and scattering amplitude}:
The two-pion spectrum with maximal isospin is shown in
Fig. \ref{fig:res2pi} together with the noninteracting energies.
The respective differences encode the two-pion scattering amplitude
for $\ell=0$ shown in Fig. \ref{fig:phase_shift} neglecting
the effect of higher even partial waves.
Its energy dependence for scattering momentum $q^2/m_\pi^2 < 1$ is described
using the effective-range expansion
\begin{align}
  q \cot \delta_0 = -\frac{1}{a_0} + \frac{r_0}{2} q^2,
	\label{eqn:ere}
\end{align}
and the scattering length $a_0$ and effective range $r_0$ are determined from
a fit using the determinant-residual
method \cite{Morningstar:2017spu} with $\mu=64$, which yields
\begin{equation}
\begin{aligned}
	m_\pi a_0 = 0.1019(88), \quad m_\pi r_0 = 9.0(2.4), \\ \chi^2/\mathrm{dof} = 1.33.
\end{aligned}
  \label{eqn:fit_results}
\end{equation}
A comparison of our scattering length with previous lattice QCD determinations
\cite{Fu:2013ffa,Sasaki:2013vxa,Beane:2005rj,Feng:2009ij,Beane:2007xs,Dudek:2012gj,Bulava:2016mks}, together with
the experimental value from \cite{Ananthanarayan:2000ht},
is shown in the supplemental material .

\begin{figure}
\includegraphics{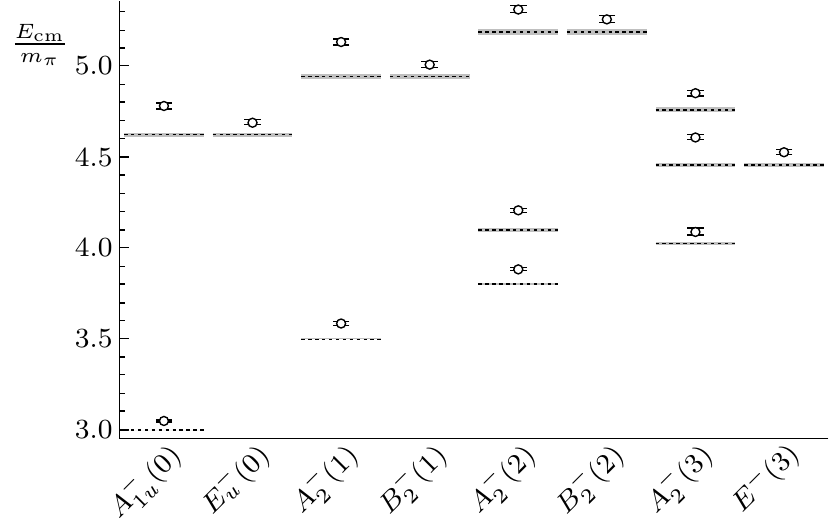}%
\caption{\label{fig:res3pi}Same as Fig. \ref{fig:res2pi} but for
the $I=3$ three-pion spectrum.}
\end{figure}

\emph{Three-pion spectrum and $1/L$ expansion}:
The three-pion spectrum with maximal isospin is shown in Fig. \ref{fig:res3pi},
displaying significant energy shifts
in all irreps.
In particular, interacting energy levels
from different irreps that contain some degeneracy of the noninteracting spectra
(e.g. $A_{1u}^-$ and $E_{u}^-$ at zero total momentum) differ substantially,
which may suggest sensitivity to different combinations of
low-energy scattering parameters.

At leading order in the $1/L$ expansion of the three-particle
quantization condition \cite{Beane:2007qr,Detmold:2008gh,Hansen:2016fzj,Sharpe:2017jej,Pang:2019dfe}
\footnote{The numerical constants appearing in this expression are evaluated in appendices A and C of \cite{Hansen:2016fzj}},
\begin{multline}
\Delta \! E_{\rm 3} =
\frac{12 \pi a_0}{m_\pi L^3} \Bigg\{
  1 - \Big(\frac{a_0}{\pi L}\Big)\mathcal{I} + \Big(\frac{a_0}{\pi L}\Big)^2(\mathcal{I}^2+\mathcal{J})
+ \frac{3\pi a_0}{m_\pi^2 L^3}
  \\
  + \frac{64 \pi^2 a_0^2 \mathcal{C}_3}{m_\pi L^3}
+ \frac{6\pi r_0 a_0^2}{L^3}
+ \Big(\frac{a_0}{\pi L}\Big)^3[
c_L \log(N_{\rm cut}) 
  -\mathcal{I}^3 
+\mathcal{I}\mathcal{J} 
  \\
+ 15 \mathcal{K}
  + \mathcal{C}_F + \mathcal{C}_4 + \mathcal{C}_5
]
\Bigg\}
- \frac{\mathcal{M}_{\rm 3, th}}{48 m_\pi^3 L^6} + O(L^{-7})
\,,
\label{eq:m3}
\end{multline}
the ground-state energy shift $\Delta \! E_\mathrm{3}$ is three times larger than the corresponding
two-particle shift $\Delta \! E_2$ \cite{Beane:2007es}.
The deviation of our numerical result
\begin{align}
	\Delta \! E_3 / \Delta \! E_2 = 2.78(21),
\end{align}
is due to two-particle effects at higher orders in $1/L$ and
the three-pion interaction.
Using the two-particle scattering length and effective range determined before,
the three-particle threshold scattering amplitude entering at $L^{-6}$ can be
isolated and we obtain
\begin{align}
	\frac{m_\pi^2 \mathcal{M}_{\rm 3, th}}{48 \, (m_\pi L)^6} = 0.0113(43).
\end{align}
While this quantity depends on how two- and three-particle effects
are separated \cite{Hansen:2016fzj,Romero-Lopez:2018rcb},
within the scheme
discussed in those references our result indicates sensitivity to three-particle
physics.

Using the nonrelativistic threshold expansion from Ref. \cite{Pang:2019dfe}
yields a result with similar significance.
From that reference, the energy of the excited states at rest are
predicted to be
\begin{equation}
  \begin{aligned}
  A_{1u}^-: \quad E_3' / m_\pi &= 4.76(8), \\
  E_u^-: \quad E_3' / m_\pi &= 4.72(2),
  \end{aligned}
\end{equation}
in good agreement with our measured values.
Further, the formalism employed in \cite{Mai:2018djl} with the scattering
parameters determined there predicts the first excited state in the
at-rest $A_{1u}^-$ irrep to be\footnote{M. Mai, private communication}
\begin{align}
	E_3' / m_\pi = 4.75
\end{align}
at our pion mass and spatial volume, which is also in agreement with
the measured value.

More work along the lines of \cite{Doring:2018xxx,Blanton:2019igq} is required
to apply the quantization condition to the energies in all irreps presented
here.
In order to facilitate further investigations in that direction,
the two-pion and three-pion spectra are made
publicly available including all correlations.
The values and covariance matrix of all extracted energies, as well as
the single-pion mass, are
given in the supplemental material,
and the bootstrap samples from this analysis are
available as ancillary files with the arXiv submission.

\section{Conclusions and outlook\label{s:conc}}
We have presented the $I=3$ three-pion spectrum in finite volume from lattice QCD
in which, for the first time, the excited states in
various irreps at zero and nonzero total momentum have been extracted.
The nonrelativistic three-particle $1/L$ expansion consistently
describes the levels in the irreps at zero total momentum.
However, the entire spectrum should be interpreted in the framework of a
full three-particle finite-volume formalism in order to corroborate
and extend our extraction of the three-pion interaction.
In the interest of facilitating those investigations, which will require generalizations
of the formulae currently available in the literature,
all spectra are made public, including their correlations.

We also described a method to reduce the computational cost
of constructing correlation functions in the stochastic variant of
distillation, achieving a speedup of more than an order of magnitude
for the set of observables considered here.
This algorithmic improvement paves the way to study
more complicated systems
such as the Roper resonance, which
has a sizable branching ratio for decays to $N\pi\pi$ as
well as $N\pi$,
and also reduces the computational cost associated with correlation
function construction
for baryon-baryon systems, which will facilitate the lattice QCD investigation
of nucleon-nucleon as well as nucleon-hyperon interactions relevant
for nuclear physics.

% If you have acknowledgments, this puts in the proper section head.
\begin{acknowledgments}
% put your acknowledgments here.
We acknowledge helpful discussions with Andria Agadjanov, John Bulava, Ken McElvain,
Daniel Mohler, Colin Morningstar,
Fernando Romero-L\'{o}pez, Akaki Rusetsky, and Andr\'{e} Walker-Loud.
We thank Christopher K\"{o}rber for useful comments on an earlier version of
the manuscript.
The authors thank the Yukawa Institute for Theoretical Physics at
Kyoto University, where part of this work was performed during the
YITP-19-01 workshop on ``Frontiers in Lattice QCD and related topics''.
The work of BH was supported by the Laboratory Directed
Research and Development Program of Lawrence Berkeley National
Laboratory under U.S. Department of Energy Contract
No. DE-AC02-05CH11231.
Calculations for this project were performed on the HPC clusters ``HIMster II'' 
at the Helmholtz-Institut Mainz and ``Mogon II'' at JGU Mainz.
Our programs use the deflated SAP+GCR solver from the openQCD
package \cite{Luscher:2011kk}, as well as the QDP++/Chroma
libraries \cite{Edwards:2004sx}.
We are grateful to our colleagues in the CLS initiative for sharing ensembles.
\end{acknowledgments}

% Create the reference section using BibTeX:
\bibliography{latticen}

% Specify following sections are appendices. Use \appendix* if there
% only one appendix.
%\appendix
%\section{}
\appendix
\section{Three-pion interpolating operators}
\label{sec:threepi}
The interpolators used in this work are given in Tables \ref{tab:int0}
to \ref{tab:int3} for zero and nonzero total momenta.
In practice, the normalization of an interpolator along each row of
the tables is arbitrary.
Correlation functions for equivalent total momenta are averaged,
and the corresponding interpolators can be obtained using
the reference rotations given in Ref.~\cite{Morningstar:2013bda} of the main text.

\begin{table}[!h]
	\caption{\label{tab:int0}Interpolating operators in the form of Eq. \eqref{eqn:threepi} with total momentum $\textbf{d}=[000]$ transforming according to row $\mu$ of irrep $\Lambda$.
Each row corresponds to one interpolator in terms of the elementals in the column heading with coefficient given in the table. Empty entries indicate a vanishing coefficient.}
\begin{ruledtabular}
	\scriptsize
\begin{tabular}{cccccccccc}
$\Lambda$ & $\mu$ &\rot{\texttt{[000][-00][+00]}} &\rot{\texttt{[000][0-0][0+0]}} &\rot{\texttt{[000][00-][00+]}} &\rot{\texttt{[000][000][000]}} &\rot{\texttt{[000][00+][00-]}} &\rot{\texttt{[000][0+0][0-0]}} &\rot{\texttt{[000][+00][-00]}} & \phantom{meh} \\
\hline
$A_{1u}^-$ & $1$ &  $1$ &  $1$ &  $1$ &           &  $1$ &  $1$ &  $1$ & \\
& &            &            &            & $1$ &            &            &            & \\
$E_u^-$ & $1$ & $-\sqrt{3}$ &  $\sqrt{3}$ &            &           &            &  $\sqrt{3}$ & $-\sqrt{3}$ & \\
 & $2$ & $1$ &  $1$ & $-2$ &           & $-2$ &  $1$ &  $1$ & \\
\end{tabular}
\end{ruledtabular}
\end{table}

\begin{table}[!h]
	\caption{Same as Table \ref{tab:int0} but for $\textbf{d}=[001]$.}
\begin{ruledtabular}
	\scriptsize
\begin{tabular}{cccccccc}
%	& &\rotatebox{-90}{\texttt{[-00][+0+][000]}} & \texttt{[0-0]} & \texttt{[000]} & \texttt{[0+0]} & \texttt{[+00]} \\
%& &  & \texttt{[0++]} & \texttt{[000]} & \texttt{[0-+]} & \texttt{[-0+]} \\
% $\Lambda$ & $\mu$ &  & \texttt{[000]} & \texttt{[00+]} & \texttt{[000]} & \texttt{[000]} \\
	$\Lambda$ & $\mu$ &\rot{\texttt{[-00][+0+][000]}} &\rot{\texttt{[0-0][0++][000]}} &\rot{\texttt{[000][000][00+]}} &\rot{\texttt{[0+0][0-+][000]}} &\rot{\texttt{[+00][-0+][000]}} & \phantom{meh} \\
\hline
$A_2^-$ & $1$ &  &   & $1$ &   & &  \\
 &&  $1$ &  $1$ &  &  $1$ &  $1$& \\
$B_2^-$ & $1$ & $1$ & $-1$ &  & $-1$ &  $1$& \\
\end{tabular}
\end{ruledtabular}
\end{table}

\begin{table}[!h]
	\caption{Same as Table \ref{tab:int0} but for $\textbf{d}=[011]$.}
\begin{ruledtabular}
	\scriptsize
\begin{tabular}{cccccccc}
$\Lambda$ & $\mu$ &\rot{\texttt{[000][-00][+++]}} &\rot{\texttt{[000][000][0++]}} &\rot{\texttt{[000][00+][0+0]}} &\rot{\texttt{[000][0+0][00+]}} &\rot{\texttt{[000][+00][-++]}} & \phantom{meh}  \\
\hline
$A_{2}^-$ & $1$ &  $1$ &           &           &           &  $1$ \\
 &     &       &           & $1$ & $1$ &            \\
 &     &       & $1$ &           &           &            \\
$B_{2}^-$ & $1$ &  $1$ &           &           &           & $-1$ \\
\end{tabular}
\end{ruledtabular}
\end{table}

\newlength{\unitlen}
\settowidth{\unitlen}{1}
\begin{table}[!h]
	\caption{\label{tab:int3}Same as Table \ref{tab:int0} but for $\textbf{d}=[111]$.}
\begin{ruledtabular}
	\scriptsize
\begin{tabular}{cccccccccccccccccc}
$\Lambda$ & $\mu$ &\rot{\texttt{[000][000][+++]}} &\rot{\texttt{[000][0+0][+0+]}} &\rot{\texttt{[000][0++][+00]}} &\rot{\texttt{[000][+00][0++]}} &\rot{\texttt{[000][+0+][0+0]}} &\rot{\texttt{[000][++0][00+]}} &\rot{\texttt{[00+][0+0][+00]}} &\rot{\texttt{[00+][+00][0+0]}} &\rot{\texttt{[00+][++0][000]}} &\rot{\texttt{[0+0][00+][+00]}} &\rot{\texttt{[0+0][+00][00+]}} &\rot{\texttt{[0+0][+0+][000]}} &\rot{\texttt{[+00][00+][0+0]}} &\rot{\texttt{[+00][0+0][00+]}} &\rot{\texttt{[+00][0++][000]}} & \phantom{meeeeeh} \\
\hline
$A_2^-$ & $1$ &  &  &  &  &  &  &  &  & $1$ &  &  & $1$ &  &  & $1$ \\
 & &  &  &  &  &  &  & $1$ & $1$ &  & $1$ & $1$ &  & $1$ & $1$ &  \\
 & & $1$ &  &  &  &  &  &  &  &  &  &  &  &  &  &  \\
 $E^-$ & $1$  &  &  & $1$ &  & $1$ & \makebox[\unitlen][c]{$-2$} &  &  &  &  &  &  &  &  &  \\
 & $2$  &  & \makebox[\unitlen][c]{$\sqrt{3}$} &  & \makebox[\unitlen][c]{$-\sqrt{3}$} &  &  &  &  &  &  &  &  &  &  &  \\
\end{tabular}
\end{ruledtabular}
\end{table}

%\clearpage
\section{Algorithm for operation-count minimization}
The relevant correlation functions are evaluated through tensor
contractions of the rank-2 meson functions defined in Eq. (5)
as shown in Eq. (6) of the main text for the single-pion correlation function.
For multi-pion correlation functions, 
linear combinations of products of these meson functions with various momenta,
but the same total momentum,
are required to compute correlation functions in a definite irrep according to the
Clebsch-Gordan coefficients discussed in the main text.
In the following we refer to an expression in terms of tensor contractions
for a given pair of source
and sink times, and each meson function with a single noise combination and
momentum, as a diagram.

The method we use to reduce the operation count required for the evaluation of
tensor contractions was proposed in the context of quantum chemistry
(Refs.~\cite{05-TR10,10.1007/11758501_39,doi:10.1021/jp9051215} in the main text) and consists
of two parts.

In the first part, for a given diagram,
using pairwise contraction of two tensors
over all joint indices as the elemental computational kernel, the list of
possible locally optimal next computational steps is determined by requiring
that the proposed step remove as many contractions as possible at the
lowest cost.
This step is irrelevant for multi-meson correlation functions as all
contractions have complexity of either $N_\mathrm{dil}^2$ if both indices
are contracted over or $N_\mathrm{dil}^3$ if only one index is contracted
with two spectator indices; this step is however useful in meson-baryon
and baryon-baryon correlation function construction where the arithmetic
operation count is reduced by powers of $N_\mathrm{dil}$ through
judicious choice of the order of elemental contractions.
This in-diagram optimization is for instance implemented in the
\texttt{einsum} function which ships with \texttt{numpy} (\url{https://github.com/dgasmith/opt_einsum}).

In the second part, each of the locally optimal proposed steps is
ranked against all diagrams to identify the step which appears most
frequently as a subexpression globally.
The best-ranking contraction is performed and the resulting intermediary
object substituted in all relevant diagrams, thereby reducing the number of
contractions required to compute the whole set of correlation functions
compared to computing diagrams individually.
This procedure is referred to as common subexpression elimination (CSE)
for instance in compiler design.

As a simple example, consider the list of two diagrams to compute shown below.
\begin{center}%
\includegraphics{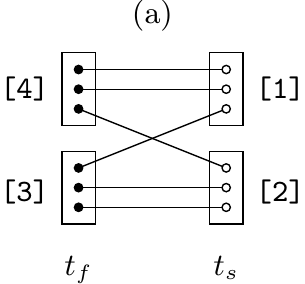}%
\hspace{2em}
\includegraphics{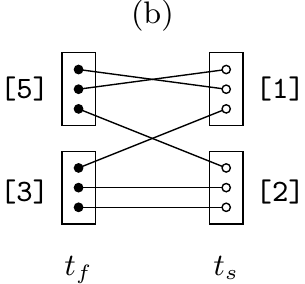}%
\end{center}%
They involve five baryon functions, i.e. rank-3 tensors,
each with a definite
momentum, irrep and noise combination labeled one to five.
The most straightforward way to evaluate those two diagrams is to combine
tensors on the source and sink times into outer products and perform the
resulting contractions in both diagrams at a cost of $2N_\mathrm{dil}^6$.

According to the first step of the algorithm discussed above,
a better way to evaluate the single diagram (a) proceeds as follows.
Viewing pairwise contraction of two tensors over all common indices
as the computational kernel, there are four possible steps:

\vspace{1em}
\noindent\mbox{%
\begin{ruledtabular}
\begin{tabular}{ccc}
	contraction step & removed indices & step complexity \\
	\hline
	\texttt{[2]} -- \texttt{[3]} & 2 & $N_\mathrm{dil}^4$ \\
	\texttt{[1]} -- \texttt{[4]} & 2 & $N_\mathrm{dil}^4$ \\
	\texttt{[1]} -- \texttt{[3]} & 1 & $N_\mathrm{dil}^5$ \\
	\texttt{[2]} -- \texttt{[4]} & 1 & $N_\mathrm{dil}^5$
\end{tabular}
\end{ruledtabular}}

\vspace{1em}%
The step complexity is given by the number of summed indices plus the number
of spectator indices.
The first and second line are the locally good choices in this
greedy algorithm, as they reduce the number of remaining contractions the most
and at the smallest cost.
Labeling the resulting rank-2 tensor from either of those as a new intermediary
\texttt{[6]} transforms diagram (a) into a diagram with contractions between
two rank-3 tensors and one rank-2 tensor remaining.
Iterating this process shows that each of the two diagrams
can be evaluated with complexity
$2 N_\mathrm{dil}^4 + N_\mathrm{dil}^2$.

The second step of the algorithm exploits the
freedom to choose between the two locally good steps in order to reduce the
global operation count.
In this example, the contraction \texttt{[2]} -- \texttt{[3]} can be
re-used in diagram (b), whereas \texttt{[1]} -- \texttt{[4]} cannot.
Therefore the first step is more beneficial, and the corresponding
subexpression, which only has to be computed once, is replaced with the intermediary \texttt{[6]} in both diagrams.

The two diagrams can thus be computed with complexity
$3 N_\mathrm{dil}^4 + 2 N_\mathrm{dil}^2$, saving one of the
computationally dominant contractions.

\begin{table}
	\caption{\label{tab:cseopt} Number of elemental
	contractions with given complexity in $N_\mathrm{dil}$ required 
	to compute all correlation functions used in this work for a fixed
	time separation on a single gauge configuration with
	and without common subexpression elimination (CSE) and
	diagram consolidation (DC).
	Diagram consolidation does not reduce the number of required
	contractions when CSE is used, but speeds up the optimization
	process.}
\begin{ruledtabular}
\begin{tabular}{l c c}
	 & $N_\mathrm{dil}^2$ & $N_\mathrm{dil}^3$ \\
	\hline
	w/o CSE, w/o DC & 36,860,400 & 44,042,400 \\
	w/o CSE, w/\phantom{o} DC & 10,035,600 & 11,810,400 \\
	w/\phantom{o} CSE & \phantom{1}2,789,370 & \phantom{11,}761,093 \\
\end{tabular}
\end{ruledtabular}
\end{table}
In order to speed up this optimization process for the large number
of diagrams required in this work, duplicate diagrams
are eliminated in a preprocessing step, which we refer to as diagram consolidation (DC).
Duplicate diagrams are produced when a given momentum combination
appears in the Clebsch-Gordan coefficients for more than one
irrep (e.g. for the $\pi(0)\pi(1)\pi(1)$ operators in
the irreps $A_{1u}^-$ and $E_{u}^-$ with zero total momentum), or in
the course of noise
averaging (e.g. for the ground-state interpolator in $A_{1u}^-$).

The number of elemental contractions required to compute all
the correlation functions used in this work are given in Table \ref{tab:cseopt}.
In total, 20,679,840 diagrams had to be evaluated, of
which 15,013,440 were consolidated before optimizing the
computation of the remaining diagrams.
After consolidation of diagrams, employing CSE reduces the operation
count by roughly a factor of 15 for the computationally dominant contractions
which scale like $N_\mathrm{dil}^3$.
%\clearpage
%\clearpage
\begin{table}[!h]
	\caption{\label{tab:r_summary} Comparison of the effective range obtained in
  this work to other lattice calculations. For convenience, the scattering length is
  shown as well. References are given in the main text.}
\begin{ruledtabular}
\begin{tabular}{ccccc}
$m_\pi \, \mathrm{[MeV]}$ & $m_\pi a_0$ & $m_\pi r_0$ & Ref. & Notes \\
\hline
200 & 0.1019(88) & 9.0(2.4) & this work & \\
396 & 0.230(19) & 12.9(3.3) & \cite{Beane:2011sc} & Fit A \\ % NPLQCD
396 & 0.226(19) & 18.1(5.3) & \cite{Beane:2011sc} & Fit B \\ %NPLQCD
396 & 0.307(13) & -0.26(13) & \cite{Dudek:2012gj} & \\ % HadSpec
282 & 0.149(10) & 21(6) & \cite{Helmes:2015gla} & $L/a=32,24,20$ \\ %ETMC
282 & 0.154(15) & 8(15) & \cite{Helmes:2015gla} & $L/a=32,24$ \\ %ETMC
233 & 0.064(12) & 18.1(8.4) & \cite{Bulava:2016mks} & \\ %Bulava, et al.
\end{tabular}
\end{ruledtabular}
\end{table}

\begin{figure}
\includegraphics{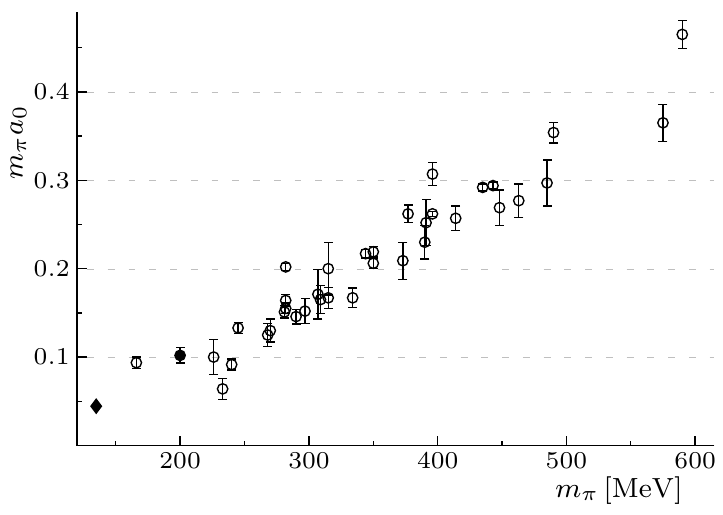}%
\caption{\label{fig:a_summary} Comparison of the scattering lengths extracted here
(filled circle) with various lattice calculations. The filled diamond corresponds
to the experimental value determined via Roy equations. The references are given in the main text.}
\end{figure}

\section{Summary of $I=2$ $\pi\pi$ scattering results from lattice QCD}
A comparison of the scattering length $a_0$ and effective range $r_0$ obtained in this work
to previous determinations from lattice QCD is shown in Figure \ref{fig:a_summary}
and Table \ref{tab:r_summary}, respectively.
These results do not account for the discretization effects inherent to the different actions used
by the various groups.
The references are given in the main text.

\section{Covariance matrix}
\label{sec:covar}
All energies and their uncertainties, as well as their normalized covariances,
for the single-pion mass, the $I=2$ two-pion spectrum and the $I=3$ three-pion
spectrum are given in Table \ref{tab:speccov}.
The single-pion mass in the first row is given in lattice units.
The scale setting for the D200 ensemble employed here is
discussed in Ref.~\cite{Bruno:2016plf} of the main text.
All other energies are quoted in units of the single-pion mass.
\begin{turnpage}
\begin{table}[!h]
	\caption{\label{tab:speccov}
	Complete spectrum and normalized covariances. Different isospin sectors
	are separated by large row spacing into $I=1$ (a single pion at rest),
	$I=2$ and $I=3$. The central values and uncertainties of center-of-mass energies $E_\mathrm{cm}/m_\pi$ in the second column are measured in units of
	the pion mass except for the single-pion energy in the first row,
	which is given in lattice units. The irrep labels are formatted as 
	in Fig. \ref{fig:res3pi}.}
\begin{ruledtabular}
	\tiny
\begin{tabular}{clcccccccccccccccccccccccccccccccccccc}
	$A_{1u}^-(0)$ & .06504(33) &$1$ \\[0.8em]
	%\hline
$A_{1g}^+(0)$ & 2.0172(17) &$-.156$ & $1$ \\
 & 3.715(14) &$-.944$ & $.146$ & $1$ \\
 & 4.897(23) &$-.920$ & $.179$ & $.873$ & $1$ \\
$E_{g}^+(0)$ & 3.624(13) &$-.974$ & $.194$ & $.922$ & $.902$ & $1$ \\
 & 4.720(21) &$-.992$ & $.147$ & $.935$ & $.915$ & $.962$ & $1$ \\
$A_{1}^+(1)$ & 2.4082(41) &$-.796$ & $.398$ & $.765$ & $.741$ & $.785$ & $.784$ & $1$ \\
 & 4.010(16) &$-.969$ & $.190$ & $.929$ & $.916$ & $.947$ & $.962$ & $.776$ & $1$ \\
 & 4.780(19) &$-.982$ & $.167$ & $.923$ & $.906$ & $.958$ & $.975$ & $.777$ & $.951$ & $1$ \\
$B_{1}^+(1)$ & 3.889(15) &$-.968$ & $.181$ & $.919$ & $.905$ & $.960$ & $.959$ & $.782$ & $.943$ & $.957$ & $1$ \\
$A_{1}^+(2)$ & 2.6280(49) &$-.837$ & $.336$ & $.805$ & $.780$ & $.831$ & $.830$ & $.725$ & $.834$ & $.833$ & $.831$ & $1$ \\
 & 2.9549(85) &$-.952$ & $.191$ & $.932$ & $.873$ & $.934$ & $.940$ & $.829$ & $.926$ & $.928$ & $.928$ & $.828$ & $1$ \\
 & 4.125(16) &$-.991$ & $.156$ & $.940$ & $.916$ & $.966$ & $.984$ & $.791$ & $.959$ & $.974$ & $.961$ & $.829$ & $.942$ & $1$ \\
$A_{2}^+(2)$ & 4.208(16) &$-.993$ & $.164$ & $.939$ & $.914$ & $.970$ & $.986$ & $.793$ & $.965$ & $.979$ & $.967$ & $.837$ & $.947$ & $.985$ & $1$ \\
$B_{1}^+(2)$ & 4.091(16) &$-.994$ & $.161$ & $.940$ & $.919$ & $.964$ & $.988$ & $.788$ & $.963$ & $.978$ & $.964$ & $.831$ & $.945$ & $.987$ & $.990$ & $1$ \\
$A_{1}^+(3)$ & 2.7868(80) &$-.664$ & $.276$ & $.628$ & $.604$ & $.661$ & $.657$ & $.556$ & $.678$ & $.660$ & $.652$ & $.635$ & $.672$ & $.657$ & $.671$ & $.665$ & $1$ \\
 & 3.344(12) &$-.915$ & $.172$ & $.896$ & $.857$ & $.893$ & $.905$ & $.769$ & $.910$ & $.904$ & $.899$ & $.808$ & $.913$ & $.909$ & $.912$ & $.906$ & $.664$ & $1$ \\
$E_{}^+(3)$ & 3.246(10) &$-.971$ & $.201$ & $.916$ & $.896$ & $.953$ & $.963$ & $.774$ & $.947$ & $.956$ & $.948$ & $.846$ & $.937$ & $.965$ & $.968$ & $.967$ & $.687$ & $.902$ & $1$ \\
$A_{1}^+(4)$ & 2.0133(29) &$-.011$ & $.310$ & $.048$ & $.059$ & $.048$ & $-.003$ & $.144$ & $.035$ & $.036$ & $.076$ & $.190$ & $.108$ & $.027$ & $.034$ & $.023$ & $.104$ & $.085$ & $.128$ & $1$ \\
 & 2.924(10) &$-.557$ & $.215$ & $.546$ & $.586$ & $.570$ & $.548$ & $.471$ & $.565$ & $.582$ & $.592$ & $.524$ & $.540$ & $.570$ & $.567$ & $.566$ & $.427$ & $.539$ & $.565$ & $.240$ & $1$ \\
 & 3.674(15) &$-.898$ & $.164$ & $.878$ & $.849$ & $.887$ & $.879$ & $.722$ & $.890$ & $.883$ & $.886$ & $.796$ & $.876$ & $.896$ & $.895$ & $.890$ & $.605$ & $.859$ & $.889$ & $.105$ & $.547$ & $1$ \\
$B_{1}^+(4)$ & 3.628(13) &$-.986$ & $.143$ & $.932$ & $.914$ & $.955$ & $.988$ & $.768$ & $.956$ & $.972$ & $.957$ & $.831$ & $.933$ & $.982$ & $.982$ & $.985$ & $.649$ & $.899$ & $.961$ & $.020$ & $.564$ & $.883$ & $1$ \\[0.8em]
%\hline
$A_{1u}^-(0)$ & 3.0478(60) &$-.164$ & $.813$ & $.164$ & $.164$ & $.182$ & $.159$ & $.462$ & $.180$ & $.155$ & $.158$ & $.268$ & $.197$ & $.160$ & $.162$ & $.168$ & $.184$ & $.172$ & $.153$ & $.102$ & $.083$ & $.149$ & $.142$ & $1$ \\
 & 4.780(17) &$-.841$ & $.130$ & $.901$ & $.772$ & $.829$ & $.828$ & $.749$ & $.831$ & $.817$ & $.832$ & $.717$ & $.877$ & $.839$ & $.843$ & $.836$ & $.561$ & $.818$ & $.815$ & $.114$ & $.516$ & $.786$ & $.826$ & $.157$ & $1$ \\
$E_{u}^-(0)$ & 4.691(15) &$-.926$ & $.305$ & $.887$ & $.854$ & $.947$ & $.915$ & $.877$ & $.898$ & $.908$ & $.911$ & $.822$ & $.930$ & $.918$ & $.921$ & $.917$ & $.640$ & $.873$ & $.909$ & $.111$ & $.540$ & $.848$ & $.905$ & $.317$ & $.846$ & $1$ \\
$A_{2}^-(1)$ & 3.5838(85) &$-.606$ & $.525$ & $.591$ & $.571$ & $.603$ & $.596$ & $.907$ & $.602$ & $.592$ & $.602$ & $.587$ & $.638$ & $.600$ & $.609$ & $.600$ & $.385$ & $.595$ & $.571$ & $.135$ & $.365$ & $.542$ & $.576$ & $.653$ & $.578$ & $.717$ & $1$ \\
 & 5.131(18) &$-.916$ & $.258$ & $.900$ & $.853$ & $.906$ & $.904$ & $.825$ & $.929$ & $.900$ & $.902$ & $.858$ & $.921$ & $.903$ & $.911$ & $.910$ & $.677$ & $.885$ & $.894$ & $.111$ & $.553$ & $.859$ & $.897$ & $.268$ & $.848$ & $.911$ & $.662$ & $1$ \\
$B_{2}^-(1)$ & 5.008(17) &$-.901$ & $.271$ & $.860$ & $.831$ & $.900$ & $.888$ & $.834$ & $.877$ & $.888$ & $.929$ & $.849$ & $.905$ & $.890$ & $.897$ & $.894$ & $.631$ & $.857$ & $.883$ & $.098$ & $.548$ & $.819$ & $.885$ & $.285$ & $.820$ & $.922$ & $.691$ & $.895$ & $1$ \\
$A_{2}^-(2)$ & 3.8840(96) &$-.661$ & $.475$ & $.654$ & $.590$ & $.672$ & $.647$ & $.684$ & $.666$ & $.649$ & $.658$ & $.871$ & $.702$ & $.650$ & $.658$ & $.656$ & $.592$ & $.661$ & $.678$ & $.229$ & $.406$ & $.620$ & $.645$ & $.443$ & $.602$ & $.721$ & $.610$ & $.751$ & $.771$ & $1$ \\
 & 4.206(11) &$-.874$ & $.298$ & $.861$ & $.807$ & $.867$ & $.860$ & $.910$ & $.856$ & $.849$ & $.852$ & $.794$ & $.937$ & $.869$ & $.872$ & $.868$ & $.611$ & $.852$ & $.861$ & $.159$ & $.507$ & $.817$ & $.850$ & $.332$ & $.852$ & $.923$ & $.764$ & $.899$ & $.878$ & $.721$ & $1$ \\
 & 5.313(19) &$-.886$ & $.233$ & $.863$ & $.814$ & $.887$ & $.872$ & $.741$ & $.878$ & $.875$ & $.872$ & $.812$ & $.892$ & $.885$ & $.886$ & $.880$ & $.815$ & $.883$ & $.899$ & $.157$ & $.530$ & $.836$ & $.866$ & $.174$ & $.799$ & $.866$ & $.540$ & $.874$ & $.832$ & $.694$ & $.827$ & $1$ \\
$B_{2}^-(2)$ & 5.258(19) &$-.895$ & $.233$ & $.844$ & $.820$ & $.887$ & $.886$ & $.759$ & $.888$ & $.887$ & $.888$ & $.810$ & $.892$ & $.885$ & $.894$ & $.899$ & $.817$ & $.874$ & $.900$ & $.128$ & $.549$ & $.823$ & $.879$ & $.191$ & $.786$ & $.876$ & $.562$ & $.874$ & $.865$ & $.703$ & $.841$ & $.915$ & $1$ \\
$A_{2}^-(3)$ & 4.088(18) &$-.416$ & $.348$ & $.376$ & $.365$ & $.420$ & $.406$ & $.445$ & $.415$ & $.415$ & $.427$ & $.430$ & $.413$ & $.398$ & $.421$ & $.417$ & $.624$ & $.399$ & $.413$ & $.121$ & $.252$ & $.356$ & $.393$ & $.351$ & $.369$ & $.453$ & $.398$ & $.475$ & $.476$ & $.457$ & $.389$ & $.492$ & $.518$ & $1$ \\
 & 4.608(14) &$-.840$ & $.253$ & $.827$ & $.769$ & $.830$ & $.826$ & $.775$ & $.818$ & $.829$ & $.831$ & $.822$ & $.881$ & $.827$ & $.833$ & $.830$ & $.646$ & $.904$ & $.834$ & $.139$ & $.519$ & $.796$ & $.819$ & $.262$ & $.790$ & $.861$ & $.628$ & $.877$ & $.870$ & $.766$ & $.845$ & $.840$ & $.834$ & $.475$ & $1$ \\
 & 4.850(17) &$-.914$ & $.176$ & $.905$ & $.840$ & $.898$ & $.903$ & $.817$ & $.891$ & $.893$ & $.892$ & $.810$ & $.967$ & $.906$ & $.913$ & $.908$ & $.639$ & $.902$ & $.900$ & $.144$ & $.533$ & $.843$ & $.895$ & $.181$ & $.867$ & $.903$ & $.626$ & $.903$ & $.879$ & $.693$ & $.927$ & $.866$ & $.862$ & $.396$ & $.870$ & $1$ \\
$E_{}^-(3)$ & 4.528(14) &$-.826$ & $.281$ & $.794$ & $.742$ & $.814$ & $.807$ & $.817$ & $.798$ & $.805$ & $.809$ & $.797$ & $.841$ & $.814$ & $.817$ & $.824$ & $.631$ & $.792$ & $.836$ & $.125$ & $.455$ & $.760$ & $.800$ & $.330$ & $.750$ & $.853$ & $.686$ & $.832$ & $.863$ & $.777$ & $.835$ & $.788$ & $.806$ & $.481$ & $.844$ & $.825$ & $1$ \\
\end{tabular}
\end{ruledtabular}
\end{table}
\end{turnpage}

\end{document}